\documentclass[12pt]{article}
\usepackage{amsfonts,amsmath,epsfig,graphicx,amssymb}
\usepackage{amsmath,epsfig,graphicx,amssymb}
\usepackage[margin=1in]{geometry}
\usepackage{amsmath}
\usepackage{amsfonts}
\usepackage{amssymb}
\usepackage{makeidx}
\usepackage{graphicx}
\usepackage{float}
\newcommand{\beq}{\begin{equation}}
\newcommand{\eeq}{\end{equation}}
\newcommand{\ben}{\begin{eqnarray}}
\newcommand{\een}{\end{eqnarray}}
\date{}
\begin{document}
\title{A New Approach to Unification}
\author{Partha Ghose \footnote{partha.ghose@gmail.com}\\Tagore Centre for Natural Sciences and Philosophy,\\ Rabindra Tirtha, New Town, Kolkata 700156, India} 
\maketitle
\begin{abstract}
This paper presents a new perspective on unifying all fundamental interactions--gravitational, electromagnetic, weak and strong--based on stochastic processes rather than conventional quantum mechanics. Earlier work by Nelson, Kac and others have established that key quantum features such as the Schr\"{o}dinger and Dirac equations together with the Born rule can be derived from classical random processes involving finite speeds and probabilistic reversals. A fundamental length scale, inherent for dimensional consistency, regularizes the infinities that typically plague conventional field theories. The method can be used to quantize electrodynamics as well as linear gravity, using the Riemann-Silberstein vector and its generalization.

To include fields beyond electromagnetism, the Riemann-Silberstein vector can be generalized to describe non-Abelian gauge fields without relying on gauge symmetry. These fields can be coupled to spin networks--geometric structures that discretize space--leading to a unified framework that includes both matter and geometrty. In the large-scale limit, the model reproduces familiar quantum field behaviour, while remaining finite and background-independent at the fundamental level. The emergence of equilibrium states resembling Wheeler-DeWill constraints in gravity adds further depth, suggesting a novel route to quantum gravity and unification grounded in physical stochasticity rather than quantization rules.

\end{abstract}

\section{Introduction}
The twentieth century has rightly been hailed as the ``quantum century.'' The majority of physicists have been trained within the dominant framework known as the Copenhagen interpretation of quantum mechanics. At its core, this view holds that an isolated microstate is fully described by a state vector \( |\psi\rangle \) evolving unitarily according to the Schr\"{o}dinger (or Dirac) equation. Yet, this framework falls short in accounting for individual outcomes in single experimental realizations. To address this gap, the postulate of measurement-induced wavefunction collapse is introduced, giving rise to a dual dynamics--unitary evolution interrupted by abrupt, non-unitary jumps.

This duality of processes has long been recognized as problematic, giving rise to paradoxes such as the double-slit experiment, Schr\"{o}dinger's cat, Wigner's friend, and the Einstein-Podolsky-Rosen (EPR) scenario. Numerous alternative interpretations have emerged---de Broglie-Bohm, many-worlds, relational quantum mechanics, QBism---but none has achieved a consensus solution~\cite{Everett, Bohm, Fuchs, rovelli-rel, peres, freire}.

The present author's view is that this lack of closure stems from two deep-rooted assumptions: (a) classical two-valued logic~\cite{Reichenbach, ghose}, and (b) the canonical Hilbert space formulation of quantum theory which assumes two-valued logic. Recent developments suggest a radically different approach in which canonical Hilbert space theory plays no role, namely that \textit{quantum phenomena may emerge from fundamentally classical stochastic processes}. Most significantly, it has been shown that the Schr\"{o}dinger equation itself, with the Born rule ingrained in it, can be \textit{derived} from Brownian motion in configuration space (Wiener processes)~\cite{nelson1, nelson2, nelson3, guerra}, while relativistic extensions involve Poisson processes with helicity-flipping events \cite{gaveau, nandi}. Observationally, the two theories are completely equivalent (both use the Schr\"{o}dinger equation and the Born rule), but they are radically different theories. In stochastic theories, the position $x$ is a random variable and not a position operator or eigenstate, and there is no fundamental quantum-classical divide as in canonical quantization--quantum processes arise when there is an imbalance between the probabilities of forward and backward drifts or between positive and negative helicity configurations. Significantly, measurement-induced collapse is a natural consequence and not an additional hypothesis in such theories \cite{pav}--it happens whenever the above imbalance disappears.

This review outlines these developments in three parts. We begin with the Nelson-Guerra-Morato program of non-relativistic stochastic mechanics, followed by the relativistic extension using Poisson-type processes. Finally, we suggest how helicity-resolved field networks may offer insights into quantum gravity.

\section{Wiener-Type Processes: Nelson's Stochastic Mechanics}
Nelson's approach \cite{nelson1, nelson2, nelson3, guerra} postulates that quantum mechanics arises from an underlying Brownian motion process. A particle trajectory $x(t)$ is modeled as a continuous but nowhere differentiable stochastic path governed by a forward and backward Fokker-Planck dynamics:
\ben
\frac{\partial}{\partial t}\rho(x,t) &=& -\frac{\partial}{\partial x}\left[b_f(x,t)\rho(x,t)\right] + \frac{\hbar}{2m}\frac{\partial^2}{\partial x^2}\rho(x,t),\\
\frac{\partial}{\partial t}\rho(x,t) &=& -\frac{\partial}{\partial x}\left[b_b(x,t)\rho(x,t)\right] - \frac{\hbar}{2m}\frac{\partial^2}{\partial x^2}\rho(x,t)
\een
where $\rho(x,t)$ is the probability density of the random variable $x(t)$. 
Adding these two equations, one gets the continuity equation
\beq
\frac{\partial}{\partial t}\rho(x,t) + \frac{\partial}{\partial x}[v(x,t) \rho(x,t)] = 0 \label{cont}
\eeq
where 
\[
v(x,t) = \frac{1}{2} \left(b_f(x,t) + b_b(x,t)\right)
\] 
is the current velocity. The difference of the forward and backward drifts, 
\[
u(x,t) = \frac{1}{2} \left(b_f(x,t) - b_b(x,t)\right)
\]
is the {\em osmotic velocity}. Subtracting the two Fokker-Planck equations one gets
\beq
u(x,t) = \frac{\hbar}{2m}\frac{\partial}{\partial x}\ln[\rho(x,t)] = \frac{\hbar}{2m}\frac{\partial_x \rho}{\rho} = \frac{\hbar}{m}\frac{\partial R}{\partial x}\label{os}  
\eeq
where $\ln\rho(x,t) = 2R(x,t)$. 

The coupled forward-backward stochastic differential equations for the position process can thus be written as
\ben
dX(t) &=& \left(v(X(t),t) + u(X(t), t)\right) + \sigma dW_f(t),\label{sde1}\\
dX(t) &=& \left(v(X(t),t) - u(X(t), t)\right) + \sigma dW_b(t).\label{sde2}
\een
where $\sigma^2$ is the diffusion coefficient.
It follows from this that the current velocity is curl-free and can be written as
\beq
v(x,t) = \frac{1}{m}\frac{\partial }{\partial x}S(x,t) \label{v}
\eeq
where $S(x,t)$ is a scalar function which can be identified with the action.  

Introducing the Lagrangian field
\beq
{{\cal{L}}} = \frac{1}{2}m (v^2 - u^2)(x,t) - V(x),
\eeq
using stochastic control theory and putting  $\hbar = m\sigma$, Guerra and Morato \cite{guerra} have derived the following differential equations for the functions $R$ and $S$:   
\ben
\frac{\partial S}{\partial t} + \frac{1}{2m}\left(\frac{\partial S}{\partial x}\right)^2 + V + V_Q &=& 0,\,\,\, V_Q = -\frac{\hbar^2}{2m}\left[\left(\frac{\partial R}{\partial x}\right)^2 + \frac{\partial^2 R}{\partial x^2}\right],\label{shj}\\
\frac{\partial R}{\partial t} + \frac{1}{2m} \left(R\frac{\partial^2 S}{\partial x^2} + 2\frac{\partial R}{\partial x}\frac{\partial S}{\partial x}\right) &=& 0 \label{con}.
\een 
The first equation is clearly a Hamilton-Jacobi equation with an additional stochastic term $V_Q$ which takes the form
\beq
V_Q = -\frac{\hbar^2}{4m}\left[\frac{\partial_x^2\rho}{\rho} -\frac{(\partial_x \rho)^2}{2\rho^2}\right]
\eeq
in terms of $\rho = e^{2R}$. The second equation, written in terms of $\rho$ and making use of eqn (\ref{v}), is
\beq
\frac{\partial \rho}{\partial t} + \partial_x\left[v\,\rho\right] =  0
\eeq
which is the continuity equation (\ref{cont}). These two coupled partial differential equations determine the stochastic process. 

By defining a complex function $\psi = \exp (R + iS/\hbar) = \sqrt{\rho}\exp (iS/\hbar)$, these two equations can be combined into the single equation
\beq
i\hbar\frac{\partial}{\partial t}\psi(x,t) = \left(-\frac{\hbar^2}{2m}\partial_x^2 + V \right)\psi(x,t)
\eeq
which is the Schr\"{o}dinger equation. 

The wave function $\psi$ therefore describes the Markov process completely:
\ben
\rho &=& |\psi|^2,\\
u &=& \frac{\hbar}{m}\partial_x \Re \ln \psi,\\
v &=& \frac{\hbar}{m}\partial_x \Im \ln \psi.
\een
This is the `Nelson map'. In other words, once the wave function is known, the probability distribution function and the current and osmotic velocities are determined. 

Pavon \cite{pav} has shown that wave function collapse occurs when the osmotic velocity $u$ vanishes, i.e. the forward and backward drifts are balanced, bridging the gap between classical and quantum processes.

\section{Poisson-Type Processes and the Dirac Equation}

Gaveau, Jacobson, Kac, and Schulman \cite{gaveau} introduced a stochastic model with a finite propagation speed $v$ based on Poisson processes with rate $a$, where a particle randomly reverses both its direction of motion and helicity. The continuum limit of such a 1D model leads to the master equation:
\[
\frac{\partial P_{\pm}}{\partial t} = -a (P_{\pm} - P_{\mp}) \mp v \frac{\partial P_{\pm}}{\partial x}
\]
for the probability densities $P_{+}$ and $P_{-}$ of the particle moving to the right and to the left respectively.
When analytic continuation is applied \cite{gaveau},
\[
c \leftrightarrow v, imc^2/\hbar \leftrightarrow a,
\]
the equation maps to the Dirac equation in the Weyl representation:
\[
i\hbar \frac{\partial\psi}{\partial t} = mc^2\sigma_x \psi + ic\hbar \sigma_z \frac{\partial \psi}{\partial x}
\]
Here, $\psi(x)$ encodes spinor amplitudes emerging from underlying directional probability amplitudes of the Poisson process. A fundamental length $\ell$, related to the Planck length $\ell_P$, regularizes the process and guarantees that $\psi$ is a probability amplitude--no Born rule is required.

This method can also be used to quantize both electrodynamics (using the Riemann-Silberstein vector $\textbf{F} = \textbf{E} + i\textbf{B}$ which is gauge invariant and helicity-resolved) and linearized gravity. In the electromagnetic case, the equation for the RS vector can be written in the Dirac form
\beq
i \hbar \partial_t \Psi = -i c \hbar \, \mathbf{\Sigma} \cdot \nabla \, \Psi + m c^2 \, \mathbf{\beta} \, \Psi.\label{bb}
\eeq
where $\Sigma_i$ are spin-1 matrices and 
\beq
\beta = \begin{pmatrix} 0 & \mathbb{I}_3\\ \mathbb{I}_3 & 0 \nonumber
\end{pmatrix}
\eeq
is a matrix that mixes the positive and negative helicity components, analogous to the Dirac $\beta$ matrix. The mass $m$ is inroduced to allow helicity flips, and the limit $m \rightarrow 0$ is taken, omitting the limit point $m=0$.

For further details, particularly regarding linearized gravity, see Nandi and Ghose \cite{nandi}.

\section{Non-abelian Vector Fields}
It now remains to be seen if non-Abelian Riemann-Silberstein fields can be constructed in a gauge-invariant manner, and whether they too can emerge fom underling stochastic processes. To that end, consider the compact Lie group $SU(N)$ with an underlying algebra $su(N)$ whose generators satisfy
\[
[T^a, T^b] = if^{abc}T^c
\]
where $a,b,c = 1,2.\cdots,N$. Define the non-Abelian RS vector by
\[
\mathbf{F_{\pm}}^a = \mathbf{E}^a \pm i\mathbf{B}^a + gf^{abc}(\mathbf{E}^b \times \mathbf{B}^c).     
\]
Then the Lagrangian is
\[
\mathcal{L} = \frac{1}{2}\text{Re} [\mathbf{F^*_{\pm}}^a \cdot \mathbf{F_{\pm}}^a] 
\]
The variation of $\mathbf{F}_{\pm}^a$ with respect to $\mathbf{E}^a$ and $\mathbf{B}^a$ gives
\begin{equation}
\delta \mathbf{F}_{\pm}^a = \delta \mathbf{E}^a \pm i \delta \mathbf{B}^a + gf^{abc} \left( \delta \mathbf{E}^b \times \mathbf{B}^c + \mathbf{E}^b \times \delta \mathbf{B}^c \right).
\end{equation}
Thus, the variation of the Lagrangian becomes
\begin{align}
\delta \mathcal{L} &= \text{Re} \left[ \left( \mathbf{F}_{\pm}^a \right)^* \cdot \delta \mathbf{F}_{\pm}^a \right] \nonumber \\
&= \text{Re} \left[ \left( \mathbf{F}_{\pm}^a \right)^* \cdot \left( \delta \mathbf{E}^a \pm i \delta \mathbf{B}^a + gf^{abc} \left( \delta \mathbf{E}^b \times \mathbf{B}^c + \mathbf{E}^b \times \delta \mathbf{B}^c \right) \right) \right].
\end{align}
Collecting terms, one obtains the field equations:
\begin{align}
\frac{\delta \mathcal{L}}{\delta \mathbf{E}^a} &= \text{Re} \left[ \left( \mathbf{F}_{\pm}^a \right)^* \right] + gf^{abc} \, \text{Re} \left[ \left( \mathbf{F}_{\pm}^b \right)^* \times \mathbf{B}^c \right] = 0, \\
\frac{\delta \mathcal{L}}{\delta \mathbf{B}^a} &= \pm \, \text{Im} \left[ \left( \mathbf{F}_{\pm}^a \right)^* \right] + gf^{abc} \, \text{Re} \left[ \left( \mathbf{F}_{\pm}^b \right)^* \times \mathbf{E}^c \right] = 0.
\end{align}
For weak coupling $g$, one can use the approximations:
\begin{equation}
\mathbf{F}_{\pm}^a \approx \mathbf{E}^a \pm i \mathbf{B}^a.
\end{equation}
Then, one gets
\begin{equation}
\text{Re} \left[ \left( \mathbf{F}_{\pm}^a \right)^* \right] \approx \mathbf{E}^a, \quad \text{Im} \left[ \left( \mathbf{F}_{\pm}^a \right)^* \right] \approx \mp \mathbf{B}^a.
\end{equation}
The field equations therefore reduce to
\begin{align}
\mathbf{E}^a +gf^{abc} \left( \mathbf{E}^b \times \mathbf{B}^c \right) &\approx 0, \\
-\mathbf{B}^a + gf^{abc} \left( \mathbf{E}^b \times \mathbf{E}^c \right) &\approx 0.
\end{align}
At leading order, one obtains the vacuum solution $\mathbf{E}^a \approx 0$, $\mathbf{B}^a \approx 0$. Including nonlinear terms,
\begin{equation}
\mathbf{B}^a \approx gf^{abc} \left( \mathbf{E}^b \times \mathbf{E}^c \right),
\end{equation}
which reflects the expected cubic non-Abelian interactions in Yang--Mills theory.

The derived field equations therefore reduce to correct Yang--Mills dynamics in the weak-field limit, suggesting that the proposed RS-based formulation is viable as a gauge-free description of non-Abelian fields.

Let us now introduce Lie-algebra valued fields
\[
\mathbf{F}_\pm = \mathbf{F}_\pm^a T^a
\]
The Lagrangian is then
\[
\mathcal{L} = \frac{1}{2} \text Re\left[ \text Tr (\mathbf{F}^*_\pm \cdot\mathbf{F}_\pm)\right]
\]
It is now possible to generalize the Biyalynicki-Birula photon wave function to the non-Abelian case by defining
 
\[
\Psi = \begin{pmatrix} \mathbf{F}_ + \\\mathbf{F}_-\end{pmatrix}^+ \label{six}\nonumber
\]
and write its wave equation in free space in the form 

\[
i \hbar \partial_t \Psi = -i c \hbar \, \mathbf{\Sigma} \cdot \nabla \, \Psi + m c^2 \, \mathbf{\beta} \, \Psi.
\]
This equation has exactly the same form as equation (\ref{bb}), and therefore emerges from underlying stochastic processes if the massive non-Abelian photons belong to the adjoint representation of $SU(N)$, it being understood that the limit $m \rightarrow 0$ (omitting the limit point $m=0$) is to be taken in the massless case.
 
\section{Spin Networks and Quantum Gravity: Blueprint}

These developments prompt us to extend stochastic methods to spin networks---the kinematic basis states of quantum geometry in Loop Quantum Gravity \cite{rovelli, smolin, ashtekar1, ashtekar1987}. In this approach, it is possible to construct a discrete stochastic framework in which helicity-resolved amplitudes propagate along spin network edges according to first-order master equations, as will be demonstrated in a forthcoming paper \cite{ghose2}. Here we mention only a few key concepts.

Let $\Gamma$ be a graph embedded in a 3-manifold representing a spin network. Each directed edge $e$ of $\Gamma$ is assigned two complex amplitudes, $\Psi^+_e(t)$ and $\Psi^-_e(t)$, representing positive and negative helicity components. The evolution equations are given by:
\begin{align}
\frac{d}{dt} \Psi^+_e &= -\lambda (\Psi^+_e - \Psi^-_e) + T^+(e), \label{eq:plus-evolution}\\
\frac{d}{dt} \Psi^-_e &= -\lambda (\Psi^-_e - \Psi^+_e) + T^-(e), \label{eq:minus-evolution}
\end{align}
where $\lambda$ is the helicity-flip rate and $T^{\pm}(e)$ are transport terms encoding adjacency-induced flow on the network. These are Gaveau-Jacobson-Kac-Schulman type master equations adapted to a discrete geometric setting \cite{gaveau}.

Defining a two-component spinor $\Psi_e = (\Psi^+_e, \Psi^-_e)^\top$ and helicity-flip operator $F_e = \sigma_1$, we can rewrite Eqs.~\eqref{eq:plus-evolution}--\eqref{eq:minus-evolution} as:
\begin{equation}
\frac{d}{dt} \Psi_e = -\lambda(I - F_e) \Psi_e + T_e,
\end{equation}
where $T_e = (T^+(e), T^-(e))^\top$.

Upon analytic continuation $t \mapsto i t$ and introduction of a fundamental length scale $\ell$, the above becomes:
\begin{equation}
i \frac{d}{dt} \Psi_e = -i \lambda \sigma_1 \Psi_e + i T_e.
\end{equation}
This form closely resembles a Dirac equation in the Weyl representation:
\begin{equation}
i \partial_t \psi = -i \sigma^i \partial_i \psi + m \psi,
\end{equation}
with spatial derivatives emerging from coarse-grained transport terms modeled as discrete gradients across adjacent nodes. 

In the absence of transport ($T^\pm(e) = 0$), the system reaches equilibrium when helicity components balance: $\Psi^+_e = \Psi^-_e$. This helicity symmetry can be written as:
\begin{equation}
(I - F_e) \Psi_e = 0.
\end{equation}
Summing over all edges gives the global constraint:
\begin{equation}
\hat{H} \Psi := \sum_e (I - F_e) \Psi_e = 0,
\end{equation}
which mimics a Wheeler-DeWitt-type condition \cite{dewitt1967, wheeler1968}, characterizing physical states as equilibrium configurations of the helicity-flipping stochastic process.

The coupling of scalar or spinor fields, defined at vertices $v$ to edge amplitudes via helicity-sensitive source terms, can also be introduced.

As will be shown, the discrete-time evolution framework admits extension to spin foam models, where each face $f$ carries helicity labels $h_f = \pm 1$. Transition amplitudes between spin network states are then expressed as:
\begin{equation}
A(s_i \rightarrow s_f) = \sum_{\{h_f\}} \prod_f w_f(h_f) \prod_v A_v(h_f \supset v),
\end{equation}
defining a helicity-resolved path integral over spin foam histories constrained by local stochastic rules. 

This formulation suggests a new probabilistic route to quantum gravity in which stochastic helicity dynamics on spin networks reproduces Dirac-type behavior and yields equilibrium conditions analogous to Wheeler-DeWitt constraints. Importantly, this approach retains background independence and facilitates direct coupling to matter, potentially bridging stochastic field quantization with loop quantum gravity.

\section{Spin Networks Realizing Full Unification: Conceptual Framework}

\subsection{Edge Labels as Non-Abelian RS Fields}

Each edge \( e \) of the spin network is labeled not only by a spin \( j \), encoding quantum geometry, but also by a Lie-algebra-valued Riemann-Silberstein (RS) field \( F^{\pm}_a(e) \in \mathfrak{su}(N) \). These are defined as:
\begin{equation}
F^{\pm}_a = E_a \pm i B_a + g f_{abc} (E_b \times B_c),
\end{equation}
where \( E_a \) and \( B_a \) are internal ``electric'' and ``magnetic'' fields, \( f_{abc} \) are the structure constants of SU(N), and \( g \) is a coupling constant. The index \( a \) encodes the gauge interaction (e.g., SU(3) for QCD, SU(2) for weak interactions, and U(1) for electromagnetism).

\subsection{Helicity-Resolved Stochastic Dynamics}

Each edge carries helicity-resolved amplitudes \( \Psi_\pm(e) \) that evolve under stochastic master equations:
\begin{equation}
\frac{d}{dt} \Psi_\pm(e) = -\lambda \Psi_\pm(e) + \lambda \Psi_\mp(e) + \mathcal{T}_\pm(e),
\end{equation}
where \( \lambda \) is a Poisson rate for helicity flipping, and \( \mathcal{T}_\pm(e) \) represents transport across the network. This dynamics mimics an internal spin-like degree of freedom undergoing random evolution.

\subsection{Intertwiners and Matter Couplings}

At each node \( v \), gauge-invariant intertwiners are introduced to enforce local SU(N) invariance. Matter fields can be introduced as vertex-based fields:
\begin{itemize}
\item Spinors \( \psi(v) \) couple via helicity projectors and gauge fields:
\[
\bar{\psi}(v) \gamma^\mu T^a \psi(v) A^a_\mu(e),
\]
\item Scalars \( \phi(v) \) couple through Yukawa-like terms:
\[
g \sum_{e \ni v} \phi(v) \Psi_\pm(e).
\]
\end{itemize}

In conventional LQG the main challenge is that it was developed primarily as a quantization of geometry, not of matter. Attempts to couple matter often require ad hoc extensions, lack a principled unification, and lead to unresolved technical problems. 

Our helicity-resolved spin network approach, which includes internal degrees of freedom and stochastic dynamics, offers a promising alternative by providing:
\begin{enumerate}

\item An integrated geometrical and internal structure (spin, helicity, gauge symmetries),

\item A natural mechanism for time evolution (via stochastic processes),

\item And a possible route to embed Dirac-type dynamics and gauge interactions consistently.
\end{enumerate}

\subsection{Continuum Limit and Emergent Unification}

Taking the continuum limit \( \ell \to 0 \), with \( \ell \sim \ell_P \) the fundamental length, the stochastic master equations yield Dirac-type equations via analytic continuation:
\begin{equation}
i \partial_t \Psi = -i \alpha^i \partial_i \Psi + \beta m \Psi + \text{gauge terms}.
\end{equation}

In this limit:
\begin{itemize}
    \item The spin \( j \)  degrees encode quantized geometry, and in semiclassical approximations or suitable limits, are expected to reproduce the Einstein field equations, consistent with the goals of canonical and covariant loop quantum gravity.
    \item The internal SU(N) indices reproduce the Standard Model gauge fields.
    \item UV divergences are tamed by the built-in cutoff \( \ell \).
    \item All fields emerge dynamically from a common discrete substrate.
\end{itemize}

Thus, this framework provides a unified description of quantum geometry, matter, and all known interactions without requiring a background spacetime.

The central premise of this approach is that a spin network, when endowed with helicity-resolved edge dynamics and constructed with internal non-Abelian structures, can provide a discrete substrate from which all known interactions-including gravity, electroweak, and strong forces-emerge in a suitable continuum limit.

Each edge of the spin network is labeled not only by a spin \( j \) from the SU(2) representation (encoding quantum geometric information) but also by additional internal degrees of freedom corresponding to gauge symmetries (e.g., SU(3), SU(2), and U(1)) as inspired by the non-Abelian generalization of the Riemann-Silberstein (RS) formalism (section 4 above). The orientation and stochastic dynamics of these edges represent helicity transitions, while vertices host intertwiners that enforce gauge and geometric constraints.

In conventional loop quantum gravity (LQG), the Ashtekar variables \((A^i_a, E^a_i)\)-where \( A^i_a \) is an SU(2) connection and \( E^a_i \) its conjugate momentum-enable a background-independent quantization of gravity. Spin networks arise as eigenstates of geometric operators like area and volume, and spin foam models define the sum-over-histories for their evolution \cite{rovelli1998, ashtekar2004}.

\paragraph{Link to Einstein Equations.} In LQG, it is not yet rigorously established that the full Einstein field equations arise in the semiclassical limit. However, semiclassical approximations using weave states or coherent states suggest that geometric expectation values reproduce classical general relativity under certain conditions \cite{thiemann, rovelli}. This correspondence is partial and still under investigation.

In the present framework, the helicity-resolved stochastic dynamics play a central role. Inspired by the work of Gaveau, Jacobson, Kac, and Schulman \cite{gaveau}, we model the time evolution of amplitudes along directed edges using first-order master equations. Upon analytic continuation and the introduction of a universal length scale (identified with the Planck length \(\ell_P\)), these equations yield Dirac-type dynamics. At equilibrium, they enforce helicity symmetry and satisfy a Wheeler-DeWitt-type constraint, effectively freezing evolution and mimicking the timeless character of canonical quantum gravity.

\paragraph{Toward Unification.} The inclusion of non-Abelian internal symmetries through a generalized RS formalism opens a pathway toward embedding the Standard Model gauge groups into the spin network structure. As each edge now carries both geometric and gauge labels, matter and gauge fields can couple locally through vertex operators, and discrete versions of Yang-Mills dynamics may be encoded in transition rules or face amplitudes in a spin foam setting.

The Planck-scale discreteness also provides a natural UV regulator, potentially taming divergences in quantum field theory and eliminating the need for traditional renormalization.

In this way, the spin network becomes a unifying scaffolding-quantizing space, encoding internal symmetries, and yielding emergent relativistic dynamics through stochastic processes in the continuum limit.

\section*{6. Concluding Remarks}

These developments suggest that while the underlying stochastic processes remain classical, quantization results from an imbalance of forward and backward processes in the nonrelativistic case, and from an imbalance of positive and negative helicities in the relativistic case. This pays two rich dividends: (a) quantization emerges without a fundamental break with the classical domain, and (b) a universal stochastic evolution generates spin-$\frac{1}{2}$, spin-1 and spin-2 fields, all sharing the same Dirac-type structure while exhibiting different transformation properties under Wigner's little groups (and internal symmetry groups like $SU(N)$ in the spin-1 case).

The framework presented here suggests a promising path toward unification of all fundamental interactions. By extending helicity-resolved stochastic dynamics to spin networks, and further embedding non-Abelian RS-type gauge structures into edge amplitudes, we encompass gravitational and gauge interactions within a single formalism. Matter fields naturally couple to helicity-labeled edges, and the presence of a fundamental length scale \(\ell\), potentially identifiable with the Planck length \(\ell\), provides a built-in UV regulator. In the continuum limit, this stochastic theory approximates standard quantum field theory, while offering an inherently background-independent and regularized description of quantum geometry. Taken together, these ingredients -- stochastic quantization, helicity dynamics, gauge coupling, and equilibrium constraints -- suggest that a full unification of matter and geometry may be achievable within this novel probabilistic framework.

Moreover, the stochastic picture offers a compelling resolution to the long-standing quantum measurement problem. Since probabilities arise from objective random processes and not from observer-induced collapse, the Born rule becomes a derived consequence rather than a postulate, and measurement is a transition between stochastic states governed by dynamical balance conditions.

The framework also provides a novel lens through which to view quantum gravity. The emergence of Dirac-type dynamics on spin networks via helicity-resolved Poisson processes leads to a Wheeler-DeWitt-like constraint as an equilibrium condition. This suggests that background-independent stochastic evolution could be a viable alternative to canonical or path-integral quantization of gravity. 

In addition, the appearance and disappearance of time--as a parameter of evolution outside of geometry--reproduces key features of canonical quantum gravity, especially the Wheeler-DeWitt equation's timeless character in equilibrium. Time emerges from the dynamics of imbalance and dissolves at equilibrium, indicating a profound connection between stochastic irreversibility and the emergence of temporal experience.

\begin{figure}[H]
    \centering
    \includegraphics[width=0.5\textwidth]{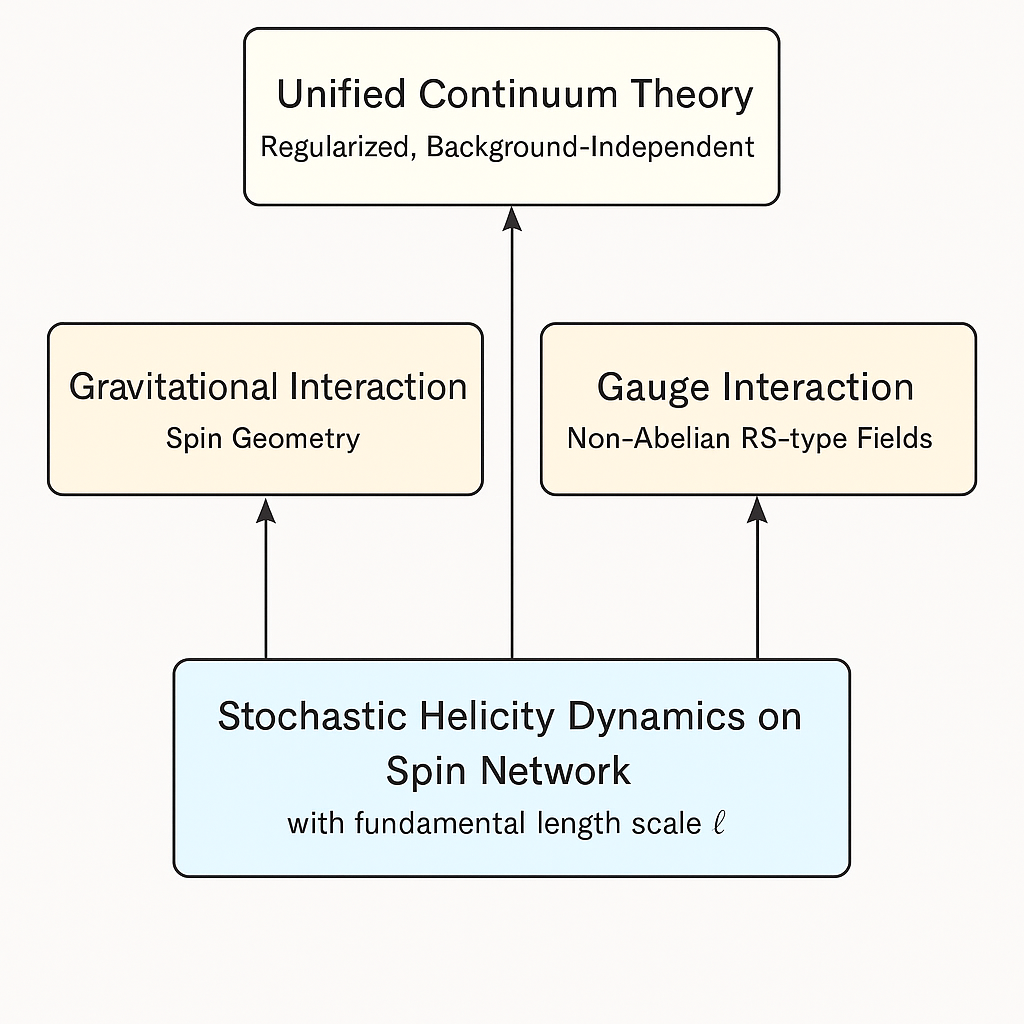}
    \caption{Flowchart illustrating the unified framework proposed in the paper. The bottom layer denotes the underlying stochastic dynamics on spin networks, governed by helicity-resolved master equations and a fundamental length scale $\ell_P$. This leads, via analytic continuation, to Dirac-type evolution equations. The gravitational sector (left) includes gravity and geometry. The gauge sector (right) accommodates non-Abelian RS structures representing the weak and strong interactions. The top row highlights equilibrium constraints such as the Wheeler-DeWitt-like condition and matter coupling mechanisms. Arrows denote conceptual and mathematical derivations linking different layers of the theory.(Flowchart prepared by ChatGPT)}
    \label{fig: unification_flow_chart}
\end{figure}

\end{document}